\documentclass[10pt]{article}
\usepackage{opex2}
\begin{document}
\title{A fourth-order Runge-Kutta in the interaction picture  method 
for  
coupled 
nonlinear Schr$\ddot{\mbox{o}}$dinger equation
} 

\author{Zhongxi Zhang, Liang Chen, Xiaoyi Bao}
\address{Department of Physics, university of Ottawa, 150 Louis Pasteur, Ottawa, K1N 6N5, Ontario, Canada \\
        }
%\address{$^1$Department of Physics, university of Ottawa, 150 Louis Pasteur, Ottawa, K1N 6N5, Ontario, Canada \\
%         $^2$BTI Photonic Systems Inc., 50 Northside Road, Ottawa, K2H 5Z6, Ontario, Canada 
%        }
\email{zzhan3@uottawa.ca}

\begin{abstract}
A fourth-order Runge-Kutta in the interaction picture (RK4IP) method is presented for 
solving the coupled
nonlinear Schr$\ddot{\mbox{o}}$dinger equation (CNLSE) that governs the light propagation in optical fibers 
with randomly varying birefringence.
%Without split-step approximation, 
The computational error of RK4IP is caused by the fourth-order Runge-Kutta algorithm,  better than the 
split-step approximation limited by the step size. 
As a result, the step size of RK4IP can have the same order of magnitude as the
dispersion length  and/or the nonlinear length of the fiber, provided 
the birefringence effect is small.
For communication fibers with random birefringence,
the step size of RK4IP can be orders of magnitude larger than the
correlation length and the beating length of the fibers, 
depending on the interaction between linear and nonlinear effects.
%As there is no need to rotate the coordinate system, 
Our approach
can be applied to the fibers having the general form of
local birefringence and treat
the Kerr nonlinearity without approximation.
%In the parameter regime where communication fibers normally operate,  
For the systems with realistic parameters,
the RK4IP results are consistent with those using Manakov-PMD approximation
\cite{Wai96,Marcuse97,Marks06}.
However, increased interaction between the linear and nonlinear terms in CNLSE leads to
increased discrepancy between RK4IP and Manakov-PMD approximation.
\end{abstract}
\vskip 0.10 in
      %(000.4430) Numerical approximation and analysis 
%(260.5430) Correlative modulation;  (060.2310) Fiber optics.}

\section{Introduction}
\label{Introduction}
%Chromatic dispersion (CD) is one of the basic factors impairing the optical performance of 
%a fiber communication system. 

In optical communication fibers,
polarization-mode dispersion (PMD, i.e., group birefringence), chromatic dispersion (CD), and Kerr nonlinearity
are the well known effects causing signal distortion.
In linear region (i. e., without Kerr nonlinearity), the signal distortion due to PMD and CD can be
predicted and compensated.
On the other hand, in the limit of pure nonlinear effect (i.e., without linear effects such as PMD and CD),
the signal amplitude
will not be affected and the pulse shape will be unchanged along the fiber.
Unfortunately both linear and nonlinear effects in real fibers are not negligible.
To accurately evaluate the signal distortion,
one must treat them simultaneously \cite{Wai96}-\cite{Shtaif06}
by solving the coupled nonlinear Schr$\ddot{\mbox{o}}$dinger equation (CNLSE),
which was proposed in 1980's  \cite{Menyuk87} 
%originally proposed in Ref. \cite{Menyuk87}
and further studied in many publications. %(cf. e.g., \cite{Wai96,Marcuse97,Marks06}).

For a non-soliton system, the CNLSE is solved with the step size determined by dispersion length $L_D$,
nonlinear length $L_N$, as well as the birefringence related parameters such as
fiber correlation length ($L_{corr}$)  (the length within which birefringence axes are randomly reoriented),
 beating length $\Lambda_{beat}$ (the length determined by birefringence strength),
and the PMD parameter $D_{PMD}$ (ps/$\sqrt{\mbox{km}}$) [1]-[3].
Ignoring the birefringence effect,  the two dimensional (2D) CNLSE can be reduced to one dimensional nonlinear
Schr$\ddot{\mbox{o}}$dinger equation
(1D NLSE), with its step size being only related with $L_D$ and $L_N$.
To efficiently solve the 2D CNLSE and 1D NLSE, various approaches have been proposed
(cf. Refs. \cite {Wai96}-\cite{Hult07} and the references therein).

%For present-day interest in communication systems,  a fundamental difficulty in efficiently solving the CNLSE
%is that the  step size is mainly restricted by the values of $\Lambda_{beat}$ (10$\sim$100 m) and $L_{corr}$
%(0.3$\sim$300 m), which are much smaller than
%$L_D$ and $L_N$ (up to hundreds of kilometers) \cite{Wai96,Marcuse97,Marks06}.
%Facing the problem of rapidly and randomly varying birefringence,
%simply decreasing the computational step size to the scale much smaller than $L_{corr}$ and $\Lambda_{beat}$ 
%not only needs  a lot of
%low efficient computation 
%\cite{Marcuse97}
%but also does not necessarily mean the improved accuracy 
%because of the computational error accumulated in all steps \cite{Hult07}.
In optical communication fibers, the values of $\Lambda_{beat}$ (10$\sim$100 m) and $L_{corr}$
(0.3$\sim$300 m) are much smaller than
$L_D$ and $L_N$ (up to hundreds of kilometers) \cite{Wai96,Marcuse97,Marks06}.
Dealing with the rapidly and randomly varying birefringence in the multispan communication fibers,
decreasing the computational step size to a very small percentage of $L_{corr}$ and $\Lambda_{beat}$ 
not only needs too much computation but also cannot ensure good enough accuracy because of the 
%is not a computationally efficient %n effective
%way. Moreover, too much computation  
%does not necessarily mean good enough accuracy, because of the 
%%associated with the  
computational error accumulated in all steps.
To efficiently solve the CNLSE, one needs an accurate approach (or algorithm) with large enough step size.  

In Ref. \cite {Marcuse97},
a  coordinate system rotating with the principal axes in each wave plate was introduced to get
the analytical solutions for the linear and nonlinear effects in CNLSE.
As a result,
the computational step size using the approach of Ref. \cite{Marcuse97} can be increased to a scale 
significantly larger than
$\Lambda_{beat}$ and $L_{corr}$. (Detailed value of its step size depends on the interaction between linear and nonlinear effects.)
Requirements for this approach  are  that
the  circular component of the local birefringence in the fiber  is negligible and that
the nonlinear Kerr effect can be approximated as its statistical average over the Poincar$\acute{\mbox{e}}$
sphere (named Manakov-PMD approximation or M-PMD approx in this work).
%Refs. \cite{Wai96, Marcuse97} also showed that
%the M-PMD approx can be used when the mixing on the Poincar$\acute{\mbox{e}}$ sphere
%is
%good enough. %, which is usually true for real transmission fibers.
Like many approaches used to solve 2D CNLSE or  1D NLSE,
another essential  %important %unnoticed 
feature of Ref. \cite{Marcuse97} is that
it needs split-step Fourier method (SSFM),
which is based on
the first order approximation of
Baker-Hausdorff formula \cite{Weiss62}
\begin{equation}
e^{A}e^{B}=e^{A+B+\frac{1}{2}[A,B]+\frac{1}{12}[A-B,[A,B]]+...},
\label{B_H_formul}
\end{equation}
where $[A, B]\equiv AB-BA$.
%Mainly 
Because of this, even in the case of zero birefringence, the step size using 
approach \cite{Marcuse97}
is restricted to  
$10\%\sim 15\%$ of $L_D$ and/or $L_N$, assuming the computational error tolerance is less than 1$\%$
of the pulse peak (cf. the results in {\bf \ref {step_size}}).
%To increase the computational step size, various improved SSFMs have been proposed.
%But they require more Fourier transformations than approach \cite{Marcuse97}.
%So far, within a given error tolerance,
%how to solve the CNLSE with maximum step size %or most efficient computation
%is still a topic under exploration.

The concept of interaction picture (IP) was
originally used in quantum mechanics. Combined with the fourth-order Runge-Kutta (RK4) algorithm,
the method of RK4 in IP (RK4IP) 
was recently proposed to study the
%dynamics of Bose-Einstein condensation and the
supercontinuum generation in optical fibers (for 1D field cases)  \cite{Hult07}.
%Unlike  various SSFMs,
Since the IP method 
does not introduce the error inherent in various SSFMs, 
the computational error of RK4IP is % mainly 
determined by the accuracy of RK4.

The approach of  Ref. \cite{Hult07} cannot be applied directly to the cases with
random birefringence, because in Ref. \cite{Hult07} the
linear dispersion operator in 1D NLSE was required to be constant along the fiber. %independ of distance.
Motivated by the analytical solution for the linear terms in CNLSE, which was proposed
in  Ref. \cite{Marcuse97} and is further discussed in the Appendix for our calculation,
we extend the RK4IP approach
from 1D NLSE % with location independent dispersion (i.e., CD)
to 2D CNLSE. %with linear dispersion (including both CD and the random birefringence).

As there is no need to rotate the coordinate system,
our approach  
can be applied to the fibers having the general form of    
local birefringence. Moreover, it treats  
the Kerr nonlinearity without approximation.
With the help of the local error method proposed in \cite{Sinkin03}, the local error of our RK4IP can be improved to
 $\sim O(h^6)$, rather than $\sim O(h^5)$ of 1D RK4IP \cite{Hult07}.
As results, for the communication fibers with random birefringence,
the step size using RK4IP can be orders of magnitude larger than
$L_{corr}$ and $\Lambda_{beat}$. %In the limit of zero  birefringence,
Without birefringence,
the step size of RK4IP can be the same order of magnitude as
$L_D$ and/or $L_N$, or, $6\sim10$ times the step size using the approach of Ref. \cite{Marcuse97}.

In the parameter regime where communication fibers operate, 
our  results are consistent  %agree %reasonably 
with those using
M-PMD approx, which was %discussed and 
used in many publications (cf. e.g., \cite{Wai96,Marcuse97,Marks06}). 
Increased interaction between the linear and nonlinear terms in CNLSE will lead to 
increased discrepancy between RK4IP and M-PMD approx. %will be increased.

%%%%%%%%%%%%%%%%%%%%%%%%%%%%%%%%%%%%

\section{From CNLSE to M-PMD approx}
\label{CNLS_Eq}
\subsection{CNLSE expressed in different forms}

To deal with the birefringence related problem, it is convenient to represent a 2D optical field 
%(in 2D Stokes space) 
with 
Dirac bra or ket notation \cite{Damask05}. 
Namely,
given a field with its $x-y$ components being $u_x$ and $u_y$, it can be denoted  as 
$|u\rangle\equiv (u_x,u_y)^T$ [or $\langle u|\equiv (u_x^*,u_y^*)$]. Thus, the CNLSE discussed in  
\cite{Wai96, Marcuse97, Marks06}    
can be written in the following form with   
retarded time $t=t_{lab}-\beta_\omega z$: 
\begin{eqnarray}
j|u\rangle_z
\!\!+\!\!\Sigma\Big[\Delta\beta |u\rangle\!+\!j\Delta\beta_{\omega}|u\rangle_t\Big]
\!-\!\frac{\beta_{\omega\omega}}{2}|u\rangle_{tt}
\!+\!\gamma\Big[\frac{5}{6}\langle u|u\rangle|u\rangle\!+\!\frac{1}{6}\sigma_{3}|u\rangle\langle u|\sigma_{3}|u\rangle
+\frac{1}{3}|v\rangle
\!\Big]\!\!=\!\!0,
\label{CNLS_3g}
\end{eqnarray}
%or
%\begin{eqnarray}
%\Big(j\frac{\partial }{\partial z}+\big[\Delta\beta \!\!+\!\!
%j\Delta\beta_{\omega}\frac{\partial }{\partial t}\big]\Sigma
%-\frac{\beta_{\omega\omega}}{2}\frac{\partial^2}{\partial t^2} \Big)
%\left(\!\!\!\!\begin{array}{c} u_x\\ u_y \end{array}\!\!\!\!\right)
%+\gamma\left(\!\!\!\!\begin{array}{c}
%(|u_x|^2+\frac{2}{3}|u_y|^2 )u_x +\frac{1}{3} u_y^2u_x^*
%\\
%(|u_y|^2+\frac{2}{3}|u_x|^2 )u_y +\frac{1}{3} u_x^2u_y^*
%\end{array}\!\!\!\!\right)
%=0
%\label{CNLS_3g2}
%\end{eqnarray}
or
\begin{eqnarray}
j|u\rangle_z
\!\!+\!\!\Sigma\Big[\Delta\beta |u\rangle\!+\!j\Delta\beta_{\omega}|u\rangle_t\Big]
\!-\!\frac{\beta_{\omega\omega}}{2}|u\rangle_{tt}
\!+\!\gamma\Big[\langle u|u\rangle|u\rangle\!-\!\frac{1}{3}\sigma_{31}|u\rangle\langle u|\sigma_{13}|u\rangle
\!\Big]\!\!=\!\!0,
\label{CNLS_3f}
\end{eqnarray}
where $|u\rangle_z\equiv \partial |u\rangle/\partial z$, $|v\rangle\equiv (u_x^*u_y^2,u_x^2u_y^*)^T$, 
and $\Delta\beta$ ($\Delta\beta_{\omega}\equiv \partial \Delta\beta/\partial \omega$) is related to  
$\Lambda_{beat}$ ($L_{corr}$) with  $\Delta\beta=\pi/\Lambda_{beat}$
($\Delta\beta_{\omega}=D_{PMD}/\sqrt{8L_{corr}}$), respectively \cite{Marcuse97}.
Here $D_{PMD}$ is the PMD coefficient
(ps/$\sqrt{\mbox{km}}$). The average DGD of a $L$-km-long fiber can be obtained using 
DGD$_{avg}=D_{PMD}\sqrt{L}$ (ps).
Obviously, the second term on the left-hand side of Eq. (\ref {CNLS_3f}) represents the
phase birefringence, while the third term relates to the group birefringence (or linear PMD).

In this work, the three components of the Pauli spin matrices are denoted as %in the standard form
\begin{eqnarray}
\sigma_1=\left(\begin{array}{cc} 0 & 1\\ 1 & 0 \end{array}\right),
\sigma_2=\left(\begin{array}{cc} 0 & -j\\ j & 0 \end{array}\right),
\sigma_3=\left(\begin{array}{cc} 1 & 0\\ 0 & -1 \end{array}\right).
\end{eqnarray}
Generally,
the birefringence-induced matrix $\Sigma$ in (\ref{CNLS_3g}) and (\ref {CNLS_3f}) has the form of
\cite{Marks06}
\begin{equation}
\Sigma \!\equiv\!
\vec \beta_0(z)\cdot\vec \sigma\!=\!
\!\left(\!\!\!\!\begin{array}{cc} \beta_{3} \!\!&\!\! \beta_1\!-\!j\beta_2\\ \beta_1\!+\!j\beta_2 \!\!&\!\!
-\beta_3\!\! \end{array}\!\!\right)
%\!=\!\sigma_3\cos 2\alpha\!+\!\sigma_1\sin2\alpha\cos2\phi\!+\!\sigma_2\sin2\alpha\sin2\phi\!
\!=\!\sigma_3\cos \theta\!+\!\sigma_1\sin\theta\cos\phi\!+\!\sigma_2\sin\theta\sin\phi\!
\label{Sigma_gener}
\end{equation}
with $\beta_{i}\equiv \langle \beta_0(z)|\sigma_i|\beta_0(z)\rangle$ ($i=1,2,3$)
and $|\beta_0(z)\rangle$ [$\vec \beta_0(z)$] being, respectively, the unit vector representing the 
fiber birefringence
in 2D Jones (3D Stokes) space \cite{Damask05}.
For the fiber with circular birefringence (e.g., spun optical fiber), 
$\beta_2\neq 0$ in (\ref {Sigma_gener}).

Parameter $\beta_{\omega\omega}=\partial^2 \beta/\partial \omega^2$ corresponds to the
CD parameter
at wavelength $\lambda$
by $\beta_{\omega\omega}$=-CD$(\lambda)\lambda^2/(2\pi c)$ (ps/nm$\cdot$km) with  c=3$\times$10$^{8}$m/s.

In Eq. (\ref{CNLS_3f}),
$\sigma_{31}=-\sigma_{13}=\sigma_3\sigma_1=\left(\begin{array}{cc} 0 & 1\\ -1 & 0 \end{array}\right )=j\sigma_2$. 
They are introduced  to   
show directly that
for a unitary transformation, e.g.,
\begin{eqnarray}
T=\left(\begin{array}{cc} t_{11} & t_{12}\\ -t_{12}^* & t_{11}^* \end{array}\right),
\;\;\;
T^{-1}=T^{\dagger}=\left(\begin{array}{cc} t_{11}^* & -t_{12}\\ t_{12}^* & t_{11} \end{array}\right),
\;\;\;
|t_{11}|^2+|t_{12}|^2=1,
\label{U_T}
\end{eqnarray}
we have $T^{\dagger}\sigma_{13}T=\sigma_{13}$, provided $T=T^*$.
This means the nonlinear term in the CNLSE is covariant for any real rotation.
%As a result, the CNLSE solution has no relation with the choice of
%the  $x-y$ directions in a
%laboratory coordinate system.
In the following sections, further  discussions on CNLSE are based on the form of (\ref {CNLS_3f}).

When the amplitude of $|u\rangle$ is viewed as the square root of optical pulse power, the 
nonlinear coefficient $\gamma$ in Eqs. (\ref {CNLS_3g})-(\ref{CNLS_3f}) relates to   
Kerr coefficient $n_2$, effective mode area $A_{eff}$, and wavenumber $k=2\pi/\lambda$
by $\gamma=n_2k/A_{eff}$. Typical values of
$\lambda=1550$ nm and
$n_2=2.6\times 10^{-20}$m$^2$/W are used throughout of the paper, unless otherwise noted.

%%%%%%%%%%%%%%%%%%%%%%%%%%%%%
%\subsection{Manakov-PMD approximation}
\subsection{M-PMD approx}
%%%%%%%%%%%%%%%%%%%%%%%%%%%%%
Eq. (\ref {CNLS_3f}) can also be expressed as
\begin{eqnarray}
\!\!\!\!\!\!\!\!j|u\rangle_z
\!\!+\!
\Sigma\!\Big[\!\Delta\beta |u\rangle\!+\!j\Delta\beta_{\omega}|u\rangle_t\!\Big]
\!-\!\frac{\beta_{\omega\omega}}{2}|u\rangle_{tt}
\!+\!\gamma\frac{8}{9}\langle u|u\rangle|u\rangle
\!\!=\!
-\frac{\gamma}{3}
\Big[\frac{1}{3}\langle u|u\rangle|u\rangle\!-\!\sigma_{31}|u\rangle\langle u|\sigma_{13}|u\rangle
\!\Big]\!,
\nonumber \!\!\!\!\!\!\!\!\!\!\!\!\!\!\!\!\!\!   \\ \!\!\!\!
\label{Manak_PMD}
\end{eqnarray}
which was named Manakov-PMD equation in Refs. \cite{Marcuse97, Marks06}.
%As well-known, the second term on the left-hand side of Eq. (\ref {Manak_PMD}) comes from the
%phase birefringence, while the third term is caused by the group birefringence (linear PMD).
The last term on the left-hand side of (\ref {Manak_PMD})
is the Kerr nonlinearity averaged over the Poincar$\acute{\mbox{e}}$
sphere with the 8/9 factor \cite{Wai96, Marcuse97, Evangelides92}.
The right-hand side of (\ref {Manak_PMD}) is the nonlinear PMD due to 
incomplete mixing on the Poincar$\acute{\mbox{e}}$
sphere \cite{Marcuse97,Marks06}.
Physically it corresponds the rapidly varying fluctuations  as the polarization state changes \cite{Marcuse97}.

In this work, the Manakov-PMD approximation (M-PMD approx) means that the variation of
the second term on the right-hand side of (\ref {Manak_PMD})
is approximated by  its statistical average over the Poincar$\acute{\mbox{e}}$
sphere [the first term on the right-hand side of (\ref {Manak_PMD})].
Thus, 
the right-hand side of (\ref {Manak_PMD}) is approximated as zero, yielding
\begin{eqnarray}
j|u\rangle_z
+\Sigma\Big[\Delta\beta |u\rangle\!+\!j\Delta\beta_{\omega}|u\rangle_t\Big]
%\!\!+\!j\!\Sigma\!\Delta\beta_{\omega}|u\rangle_t
\!-\!\frac{\beta_{\omega\omega}}{2}|u\rangle_{tt}
\!+\!\gamma\frac{8}{9}\langle u|u\rangle|u\rangle
\!\!=0.
\label{Manak_PMD_App}
\end{eqnarray}
Without linear birefringence, M-PMD approx (\ref {Manak_PMD_App}) reduces to Manakov equation \cite{Wai96, Marcuse97, Evangelides92}.

The form of Eq. (\ref {Manak_PMD_App}) is different from, 
but equivalent to, 
the M-PMD approx proposed in
Refs. \cite{Wai96,Marcuse97,Marks06}.
In fact one can obtain the published M-PMD approx, e.g.,
Eq. (68) of Ref. \cite{Marks06},
by substituting the unitary transformations $|u\rangle=RT|\tilde \Psi\rangle$
into (\ref {Manak_PMD_App}), with $T$ being  given by (\ref {U_T}) and  
$R=\left(\begin{array}{cc} \cos\alpha  & -\sin\alpha\\ \sin\alpha & \cos\alpha \end{array}\right)$, yielding
$R_z=\alpha_z\left(\begin{array}{cc} -\sin\alpha  & -\cos\alpha\\ \cos\alpha & -\sin\alpha \end{array}\right)$,
$jR^{\dagger}R_z=\alpha_z\sigma_2$, 
and $[\sigma_2\alpha_z +\sigma_3\Delta \beta]T+jT_z=0$.

\section{RK4IP: extension from 1D to 2D}
\label{RK4IP} 
 
\subsection{RK4IP solution for 1D NLSE}
\label{RK4IP_1d}

Without birefringence, Eq. (\ref{CNLS_3f}) can be reduced to 
1D NLSE  
\begin{equation}
j\frac{\partial u}{\partial z}-\frac{\beta_{\omega\omega}}{2}\frac{\partial^2 u}{\partial t^2}+\gamma|u|^2u=0,
\label{NLSE_1d} 
\end{equation}
which can be formally viewed as  
\begin{equation}
j\frac{\partial u}{\partial z}+(\hat D +\hat N)u=0,\;\;\; \hat D=-\frac{\beta_{\omega\omega}}{2}\frac{\partial^2 }{\partial t^2},
\;\;\;\hat N=\gamma|u|^2. 
\label{NLSE_1d_op}
\end{equation}
Introducing transformation $u=\exp[j(z-z_0)\hat D]u^I$, NLSE in the IP has the form 
\cite{Hult07}
\begin{equation}
j\frac{\partial u^I}{\partial z}+\hat N^Iu^I=0
\label{NLSE_1d_IP}
\end{equation}
or
\begin{equation}
\frac{\partial u^I}{\partial z}=j\hat N^Iu^I\equiv f(z, u^I)
\label{NLSE_1d_IP_2}
\end{equation}
with 
\begin{equation}
\hat N^I=\exp[-j(z-z_0)\hat D]\hat N\exp[j(z-z_0)\hat D]\equiv \hat {I}_{D}^{\dagger}\hat N\hat {I}_{D}
\label{N_IP}
\end{equation}
the nonlinear operator represented in the IP. % Note that in (\ref {N_IP}), opterator $\hat D$  
Given $u(z_n,t)$, the next step field $u(z_{n+1},t)$ ($z_{n+1}=z_n+h$) governed by  
Eq. (\ref {NLSE_1d_IP}) can be obtained using RK4, 
with the local error of $O(h^5)$. Choosing $z_0=z_n+h/2$ can reduce
the required FFTs by $50\%$ (compared to the choice of $z_0=z_n$), yielding \cite{Hult07} 
\begin{eqnarray}
\!\!\!\!\!\!\!&\!\!\!\!\!\!\!&\!\!\!\!\!\!\!
u(z_n+h,t)=\exp[j\frac{h}{2}\hat D]\big[u_n^I+k_1h/6+k_2h/3+k_3h/3\big]+k_4h/6
\nonumber\\
\!\!\!\!\!\!\!&\!\!\!\!\!\!\!&\!\!\!\!\!\!\!
u_n^I=\exp[j\frac{h}{2}\hat D]u(z_n,t)
\nonumber\\
\!\!\!\!\!\!\!&\!\!\!\!\!\!\!&\!\!\!\!\!\!\!
k_1=f(z_n,u_n^I)=j\exp[j\frac{h}{2}\hat D]\hat N\big(u(z_n,t)\big)u(z_n,t)
\nonumber\\
\!\!\!\!\!\!\!&\!\!\!\!\!\!\!&\!\!\!\!\!\!\!
k_2=f(z_n\!+\!\frac{h}{2},u_n^I\!+\!\frac{h}{2}k_1)=j\hat N\big(u_n^I+\frac{h}{2}k_1\big)[u_n^I+\frac{h}{2}k_1]
\nonumber\\
\!\!\!\!\!\!\!&\!\!\!\!\!\!\!&\!\!\!\!\!\!\!
k_3=f(z_n\!+\!\frac{h}{2},u_n^I\!+\!\frac{h}{2}k_2)=j\hat N\big(u_n^I+\frac{h}{2}k_2\big)[u_n^I+\frac{h}{2}k_2]
\nonumber\\
\!\!\!\!\!\!\!&\!\!\!\!\!\!\!&\!\!\!\!\!\!\!
k_4\!=\!\exp[-j\frac{h}{2}\hat D]f(z_n\!\!+\!\!h,u_n^I\!\!+\!\!hk_3)
\!\!=\!\!j\hat N\!\Big(\!\exp[j\frac{h}{2}\hat D][u_n^I\!+\!hk_3]\!\Big)\! \exp[j\frac{h}{2}\hat D][u_n^I\!+\!hk_3]
\label{RK4IP_1d}
\end{eqnarray}

Note that in (\ref {N_IP}) and (\ref {RK4IP_1d}), the linear operator $\hat D$ for 1D case
was assumed to be  unchanged  in $z$ direction. 
For a fiber with random birefringence, the transformation  
$\hat I_{D}(z,z_0)\equiv\exp[j(z-z_0)\hat D]=\exp[-j(z-z_0)\beta_{\omega\omega}(\partial^2/\partial t^2)/2]$ 
%which was
%used to obtain the nonlinear operator $\hat {N}_I$ 
in (\ref {N_IP}) needs to be modified correspondly, 
%becomes quite complicated,  
%(cf. detailed  below.  
which is the key point to extend the RK4IP of \cite{Hult07} to 2D case.

%%%%%%%%%%%%%%%%%%%%%%%%%%%%%%%%%%%%%%%%%%%%%%%%%%%%%%%%%%%%%%%%%%%%%%%%%%
%%%%%%%%%%%
\subsection{RK4IP solution for 2D CNLSE with random birefringence}
%%%%%%%%%%%%%
%%%%%%%%%%%%%%%%%%%%%%%%%%%%%%%%%%%%%%%%%%%%%%%%%%%%%%%%%%%%%%%%%%%%%%%%%%%
Eq. (\ref {CNLS_3f}) can also be viewed as 
\begin{eqnarray}
\!\!\!\!\!\!&\!\!\!\!\!&\!\!\!\!\!\!\!
j\frac{\partial |u\rangle}{\partial z}+(\hat {D}_2 +\hat N_2)|u\rangle=0,\;\;\; 
\nonumber \\
\!\!\!\!\!&\!\!\!\!\!&\!\!\!\!\!\!
\hat {D}_2=
\vec \beta_0(z)\!\cdot\!\vec \sigma\big(\Delta\beta \!+\!j\Delta\beta_{\omega}\frac{\partial }{\partial t}\big) 
\!-\!\frac{\beta_{\omega\omega}}{2}\frac{\partial ^2}{\partial t^2},
\;\;\;\hat N_2=
\gamma\big[\langle u|u\rangle\!+\!\frac{1}{3}\sigma_{13}\langle u|\sigma_{13}|u\rangle\big].
\label{CNLS_op}
\end{eqnarray}
Introducing the transformation $|u_n\rangle=\hat I(z_n,z_0)|u_n^I\rangle$, with the operator 
$\hat I(z,z_0)$ given by
(\ref {trans_t}), the CNLSE (\ref {CNLS_op}) can be expressed in IP as %\cite{Hult07}
\begin{equation}
\frac{\partial |u^I(z)\rangle}{\partial z}=j\hat N_2^I(z)|u^I(z)\rangle\equiv |f(z, u^I)\rangle,\;\;\; 
\hat {N}_2^I(z)=\hat I^\dagger(z,z_0)\hat {N}_2(z)\hat I(z,z_0).
\label{CNLS_IP}
\end{equation}

Given $|u_n\rangle$ (the 2D field at $z_n$), the next step field $|u_{n+1}\rangle$
can be obtained from Eq. (\ref {CNLS_IP}) by RK4.  
As in 1D case of Ref. \cite{Hult07}, one can 
choose $z_0=z_n+h/2$ to minimize the required number of FFTs, which leads to      
\begin{eqnarray}
\!\!\!\!\!\!\!\!&\!\!\!\!\!\!\!\!&\!\!\!\!\!\!\!\!\!
|u_{n+1}\rangle\!=\!\hat I(z_{n+1},z_0) |u_{n+1}^I\rangle
%\!=\!e^{\int_{z_n+h/2}^{z_n+h}\hat D(z')dz'}
=\hat{d}(\frac{h}{2})\hat{M}_2\Big[|u_n^I\rangle+|k_1\rangle\frac{h}{6}+|k_2\rangle\frac{h}{3}+|k_3\rangle\frac{h}{3}\Big]+|k_4\rangle\frac{h}{6}
\nonumber\\
\!\!\!\!\!\!\!\!&\!\!\!\!\!\!\!\!&\!\!\!\!\!\!\!\!\!
|u_n^I\rangle=\hat I^\dagger(z_n,z_0)|u_n\rangle=\hat{d}^{\dagger}(-\frac{h}{2})\hat{M_L}^\dagger|u_n\rangle
=\hat{d}(\frac{h}{2})\hat{M_1}|u_n\rangle
%=e^{-\int_{z_n+h/2}^{z_n}\hat D(z')dz'}|u_n\rangle
\nonumber\\
\!\!\!\!\!\!\!\!&\!\!\!\!\!\!\!\!&\!\!\!\!\!\!\!\!\!
|k_1\rangle=|f(z_n,u_n^I)\rangle
\nonumber\\
\!\!\!\!\!\!\!\!&\!\!\!\!\!\!\!\!&\;
=j\gamma\hat{d}(\frac{h}{2})\hat{M_1}
%j\gamma e^{-\int_{z_n+h/2}^{z_n}\hat D(z')dz'} 
\Big[\langle u_n|u_n\rangle+\frac{\sigma_{13}}{3}\langle u_n|\sigma_{13}|u_n\rangle\Big] |u_n\rangle
\nonumber\\
\!\!\!\!\!\!\!\!&\!\!\!\!\!\!\!\!&\!\!\!\!\!\!\!\!\!
|k_2\rangle=|f(z_n+\frac{h}{2},u_n^I+\frac{h}{2}k_1)\rangle
\nonumber\\
\!\!\!\!\!\!\!\!&\!\!\!\!\!\!\!\!&\;
=j\gamma \Big[\langle u_n+\frac{hk_1}{2}|u_n+\frac{hk_1}{2}\rangle
+\frac{\sigma_{13}}{3}\langle u_n^I+\frac{hk_1}{2}|\sigma_{13}|u_n^I+\frac{hk_1}{2}\rangle\Big] |u_n^I+\frac{hk_1}{2}\rangle 
\nonumber\\
\!\!\!\!\!\!\!\!&\!\!\!\!\!\!\!\!&\!\!\!\!\!\!\!\!\!
|k_3\rangle=|f(z_n+\frac{h}{2},u_n^I+\frac{h}{2}k_2)\rangle
\nonumber\\
\!\!\!\!\!\!\!\!&\!\!\!\!\!\!\!\!&\;
=j\gamma \Big[\langle u_n+\frac{hk_2}{2}|u_n+\frac{hk_2}{2}\rangle
+\frac{\sigma_{13}}{3}\langle u_n^I+\frac{hk_2}{2}|\sigma_{13}|u_n^I+\frac{hk_2}{2}\rangle\Big] |u_n^I+\frac{hk_2}{2}\rangle
\nonumber\\
\nonumber\\
\!\!\!\!\!\!\!\!&\!\!\!\!\!\!\!\!&\!\!\!\!\!\!\!\!\!
|k_4\rangle=\hat{d}(\frac{h}{2})^\dagger\hat{M}_2^\dagger|f(z_n+h,u_n^I+hk_3)\rangle
\nonumber\\
\!\!\!\!\!\!\!\!&\!\!\!\!\!\!\!\!&\;
=j\gamma \Big[\langle v_n|v_n\rangle
+\frac{\sigma_{13}}{3}\langle v_n|\sigma_{13}|v_n\rangle\Big] |v_n\rangle
,\;\;\;\;
%\nonumber\\
%\!\!\!\!\!\!\!\!&\!\!\!\!\!\!\!\!&\!\!\!\!\!\!\!\!\!
|v_n\rangle=\hat{d}(\frac{h}{2})\hat{M}_2|u_n^I+hk_3\rangle,
%e^{\int_{z_n+h/2}^{z_n+h}\hat D(z')dz'}|u_n^I+hk_3\rangle,
\label{RK4IP_2d}
\end{eqnarray}
where $|u_n^I+ck_i\rangle\equiv |u_n^I\rangle+c|k_i\rangle$ $(i=1,2,3, \; c=\frac{h}{2}, h)$,
\begin{eqnarray}
\!\!\!\!\!\!\!&\!\!\!\!\!\!\!&\!\!\!\!\!\!\!
\hat{M}_1\equiv\hat{m}_{N_{h/2}}\hat{m}_{N_{h/2}-1}\cdots\hat{m}_{2}\cdot\hat{m}_{1}
,\;\;\;\;\;\;
\hat{M}_2\equiv\hat{m}_{N_h}\cdot\hat{m}_{N_h-1}\cdots\hat{m}_{N_{h/2}+2}\cdot\hat{m}_{N_{h/2}+1},
\nonumber \\
\!\!\!\!\!\!\!&\!\!\!\!\!\!\!&\!\!\!\!\!\!\!
\hat{M}_L\equiv\hat{m}_1(-)\cdot\hat{m}_2(-)\cdots\hat{m}_{N_{h/2}}(-)
=\hat{m}_1^\dagger\cdot\hat{m}_2^\dagger\cdots\hat{m}_{N_{h/2}}^\dagger,
\label{M12}
\end{eqnarray}
and $\hat d(h/2)$ can be obtained  according to (\ref {trans_t}).
Introduced in (\ref {trans_t}), $N_{h}$ in (\ref {M12}) is the index of the last plate, whereas 
the unitary matrix $\hat{m}_i$ in (\ref {M12}) and its Hermitian (i.e., its self-adjoint matrix) 
$\hat{m}_i^\dagger$ satisfy    
$\hat{m}_i^\dagger\!=\hat{m}_i^{-1}$.
To order all $\hat{m}_i$ in one direction, the indexes in $\hat{M}_L$
[given in (\ref {M12})] does not follow the ordering 
rule introduced in the Appendix.
The minus sign in $\hat{m}_i(-)$ ($i=1,2...$) is used to
denote 
$\delta_i<0$, where $|\delta_i|$ is the length of the $i$th plate discussed in the Appendix. 
Thus we have
$\hat{m}_i(-)=\hat{m}_i^\dagger$.
Obviously, when $\gamma=0$, (\ref {RK4IP_2d}) yields   
$|u_{n+1}\rangle\!=\hat {d}(\frac{h}{2})\hat {M}_2 \hat{d}(\frac{h}{2})\hat M_1|u_{n}\rangle$, which
is consistent with (\ref {trans_t}) [or (\ref {trans}) and (\ref {M_solutn})] given in the Appendix.
%%%%%%%%%%%%%%%%%%%%%%%%%%%%%%%%%%%%
%%%%%%%%%%%%%%%%%%%%%%%%%%%%%%%%%%%
%%%%%%%%%%%%%%%%%%%%%%%%%%%%
\section{Applications and discussions}
%%%%%%%%%%%%%%%%%%%%%%%%%
In the following numerical calculations, 
the input optical field
$u_{\mathrm{in}}(t)\equiv |\vec {u}_{\mathrm{in}}(t)|$ is assumed to be a periodic repetition
of $N$-bit (N=16) de Bruijn sequence, i.e.,
$u_{\mathrm{in}}(t)=\sum_{n=-\infty}^{\infty} d_B(t-nNT_{\mathrm{b}})$, where 
 $d_B(t)=\sum_{i=0}^{N-1}a_ip(t-iT_{\mathrm{b}})$ and
$T_{\mathrm{b}}$ is the time interval of each bit.
Here $p(t)$ determines the elementary input pulse shape and $a_i$ is the logic value of the $i$th bit.
Within the time interval [0, $T_{\mathrm{b}}$], the elementary forms of RZ and NRZ pulses (or Marks)
are assumed to be
$p(t)\!=\!\sqrt{2E_{\mathrm{b}}/T_{\mathrm{b}}}\cos[\frac{\pi}{2}\cos^2(\frac{\pi t}{T_{\mathrm{b}}})]$ 
and $p(t)\!=\!\sqrt{E_{\mathrm{b}}/T_{\mathrm{b}}}$,
respectively, with $E_{\mathrm{b}}$  the optical energy per transmitted bit \cite{Forestieri00,Wang04}.
Outside this time interval, $p(t)$ is zero.  %In the following discussion, 
Obviously, $E_b/T_b$ is the mark power.
To give the NRZ optical pulses slightly rounded edges, the input pulses are generated by passing through 
an input optical filter \cite{Marcuse97}. In this work, it is assumed to be fifth-order 
Bessel type  with bandwidth $B$.

%The grid points of $2^{10]$%The FFT with for each grid points 
In all simulations, the fiber loss has been neglected, implying that there are no optical amplifiers 
in the system and, consequently, no amplifier (ASE) noise.

Only OOK format is considered in this section. Before photodetection, the optical 
signal is assumed to be filtered by a Fabry-P$\acute{\mbox{e}}$rot type channel filter with 
bandwidth $B_o=6.9/T_b$ \cite{Wang04, Wiesenfeld05}.   
The square-law-detected signal is then
electrically filtered by a fifth-order Bessel filter with 
bandwidth $B_e=0.8/T_b$ \cite{Marks06, Wiesenfeld05, Ibragimov00}.
In the following figures,  the electrically filtered photoelectric pulses are represented in  
units of W, %for comparison with the optical pulse,  
since detected current corresponds to optical 
power \cite{Marcuse97}.

In our computation, the approach of adaptive step size is used. Namely,
given step size $h$, one can use RK4IP (\ref {RK4IP_2d}) to obtain 
$|u(x_n+h/2)\rangle_f$ (the fine solution at $x_n+h/2$) 
and $|u(x_n+h)\rangle_c$ (the coarse solution at $x_n+h$)
from $|u(x_n)\rangle_{ob}$ (the obtained solution at $x_n$).
Similarly, based on $|u(x_n+h/2)\rangle_f$, the second fine solution $|u(x_n+h)\rangle_f$ 
can be obtained.
Then the next step size can be 
%increased (decreased), respectively, if the difference between 
adjusted, depending on the difference between
$|u(x_n+h)\rangle_f$ and $|u(x_n+h)\rangle_c$.
% is smaller (larger) than the low (up) boundary of the 
%local
%error.
According to Ref.\cite{Sinkin03}, the local error of RK4IP can be improved %decreased 
from $O(h^5)$ \cite{Hult07} 
to $\sim O(h^6)$
by using 
$|u(x_n+h)\rangle_{ob}=[16|u(x_n+h)\rangle_f-|u(x_n+h)\rangle_c]/15$.

%%%%%%%%%%%%%%%%%%%%%%%%%%%
%\subsection{RK4IP and SSFM results in the cases of zero birefringence}
\subsection{RK4IP and SSFM comparison: zero  birefringence}
\label{RK4IP_SSFM}
%%%%%%%%%%%%%%%%%%%%%%%%i
\begin{figure}
\vskip -6mm
\begin{center}
    \begin{tabular}{cc}
     %\resizebox{105mm}{!}{\includegraphics{RK4IP_SSFM.eps}}%CNLSE no biref
     \resizebox{105mm}{!}{\includegraphics{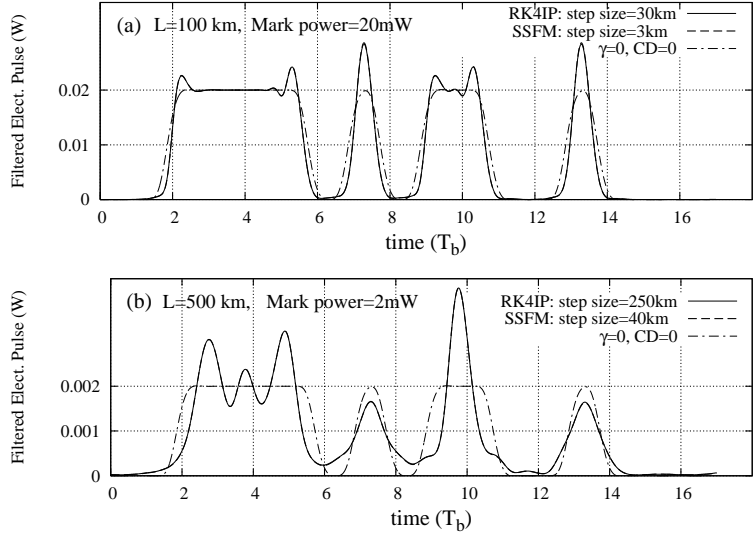}}%CNLSE no biref
     \end{tabular}
    \vskip  -2mm
\caption{Electrically filtered pulse as a function of time ($T_b=200$ps),
             with CD=17 ps/nm$\cdot$km and fiber birefringence being neglected.
             The nonlinear coefficient $\gamma=1.26$/(W$\cdot$km)
             is obtained using %wavelength $\lambda=1550$ nm,
             effective mode area
             $A_{eff}=80$ $\mu$m$^2$. 
             The input mark power $E_b/T_b$ is 20 mW for L=100 km (a) and 2 mW for 
             L=500 km (b). 
             There is no visible difference between  RK4IP solution  
             (\ref {RK4IP_2d}) and SSFM solution 
             (\ref {SSFM_1D}).
            }
    \label{fig_RK4IP_SSFM}
\end{center}
\end{figure}

%%%%%%%%%%%%%%%%
Without birefringence,  2D CNLSE (\ref {CNLS_3f}) can be reduced to 1D NLSE (\ref {NLSE_1d})
%\begin{equation}
%j\frac{\partial u}{\partial z}-\frac{\beta_{\omega\omega}}{2}\frac{\partial^2 u}{\partial t^2}+\gamma|u|^2u=0,
%\label{NLSE_1d}
%\end{equation}
with its split-step solution being ($h=z_{n+1}-z_n$)
\begin{equation}
u(z_{n+1},t)\approx F^{-1}[e^{j\frac{\beta_{\omega\omega}}{2}\omega^2h}\tilde {U}],
\;\; \tilde {u}= u(z_{n},t)e^{j\gamma |u(z_n)|^2 h},
\label{SSFM_1D}
\end{equation}
where $\tilde {U}$ is the Fourier transformation of $\tilde {u}$, while
$F^{-1}$ denotes inverse Fourier transformation.
Note that, in the case of zero birefringence, the SSFM result of 
Ref. \cite{Marcuse97} can be reduced to (\ref {SSFM_1D}).  
 
Obviously, (\ref {SSFM_1D}) 
should yield the same result as the RK4IP solution (\ref {RK4IP_2d}) (without  birefringence).
This is numerically confirmed by considering a NRZ-OOK pulse train launched into the fiber with
CD=17 ps/(nm$\cdot$km)
and $A_{eff}$=80$\mu$m$^2$ [$\gamma$=1.26 (W$\cdot$km)$^{-1}$]. The bandwidths of the electrical filter and the input optical filter are
$B_e=4$G and $B=1.75B_e$. The input mark power is 20 mW for L=100 km (a) and 
2 mW for L=500 km (b). 
As shown in
Fig. \ref {fig_RK4IP_SSFM},  the results of RK4IP without birefringence agrees very well with  
those using SSFM (\ref {SSFM_1D}).  
%%%%%%%%%%%%%%%%%%%%%%
\subsection{Step size of RK4IP}
\label{step_size}
%%%%%%%%%%%%%%%%%%%%%%%%%%
Here, the step size means that, within given computational
error tolerance ($<1\%$ of the pulse peak), the  maximum allowable step size at the end of the fiber. 
\begin{center}{\footnotesize Table 1.
Step size $\Delta z$ using RK4IP and SSFM
of Ref. \cite {Marcuse97} obtained %associated
with given
L$_N$ and
fiber length L.
The dispersion length L$_D\sim240$ km %(cf. Fig. \ref{fig_RK4IP_SSFM})
and the computational accuracy $<1\%$ of the pulse peak.
              }
\end{center}
\vskip -4mm
\begin{table}
\begin{center}
\begin{tabular}{|c|c|c c|}
\hline
\multicolumn{1}{|c}{}
&\multicolumn{1}{|c}{random birefringence}
&\multicolumn{1}{|c}{$\;\;$no$\!$}
&\multicolumn{1}{l|}{$\!\!\!\!\!\!\!\!$birefringence$\;\;\;\;\;\;\;\;$}
\\
                            &  RK4IP   & RK4IP  & SSFM of \cite {Marcuse97}
\\ \hline
$\Delta z$ (L$_N\sim 40$km, L=100 km)  &  7  km    & 30 km  &  3   km
\\ \hline
$\Delta z$ (L$_N\sim 400$km, L=500 km) &  45 km    & 250km  &  40  km
\\ \hline
\end{tabular}
\end{center}
\end{table}

Table 1 shows the step sizes using RK4IP 
for the fibers of Fig. \ref {fig_RK4IP_SSFM} (a) and (b)
with random birefringence 
[$\Lambda_{beat}$ =50m, $L_{corr}$=10m, and $D_{PMD}$=1.0ps/(km)$^{1/2}$, yielding average DGD$\approx$22ps],
compared with the step size using RK4IP (without birefringence) and the step size using (\ref {SSFM_1D}), which is  
the SSFM of Ref. \cite{Marcuse97} in the case of  
zero birefringence.
Notice that the fiber dispersion length
$L_D\sim 1/(\beta_{\omega\omega}\Delta f^2)$ ($\Delta f$ is the signal bandwidth)
for the two cases of Fig. \ref {fig_RK4IP_SSFM} is $\sim 240$ km, while
the nonlinear length $L_N\sim 1/(\gamma P)$ 
is $\sim$ 40 km for the case of Fig. \ref {fig_RK4IP_SSFM} (a) and 400 km for (b). 

In each case of zero birefringence, the RK4IP step size is around the smaller one 
of  L$_D$ and L$_N$, or, about 6$\sim$10 times of  the step size of SSFM of Ref. \cite{Marcuse97}, 
which is around $10\%\sim15\%$ of L$_D$ and/or L$_N$. 

Taking into account random birefringence, the RK4IP step size is decreased from 30km to 7 km for the case of 
L$_N\sim40$ km (L=100km) 
and from 250 km to 45km for L$_N\sim400$ km (L=500km).
In general, the stronger the interaction between the linear and nonlinear parts in CNLSE,
the more impact of  $L_{corr}$ and $\Lambda_{beat}$
on the step size, assuming that $L_D$ and $L_N$ are much larger than the former.
(For the same reason, the step size of approach \cite{Marcuse97} also needs to be decreased correspondingly.)

%%%%%%%%%%%%%%%%%%%%
\subsection{RK4IP solutions and M-PMD approximations with random birefringence}

%%%%%%%%%%%%%%%%%%%%
%which is noticeable different from the dot-dashed and solid curves.

\begin{figure}
\vskip -6mm
\begin{center}
    \begin{tabular}{cc}
     %\resizebox{110mm}{!}{\includegraphics{MnkvCNLSE_NRZ_OOK5spn113P10CD17t2.eps}}%CNLSE/CNLSE_2
     \resizebox{110mm}{!}{\includegraphics{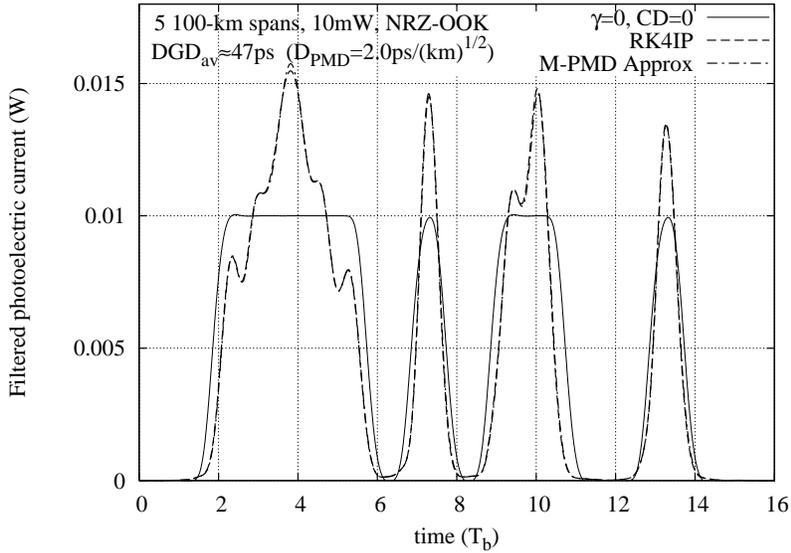}}%CNLSE/CNLSE_2
     \end{tabular}
    \vskip  -2mm
    \caption{The filtered photoelectric current (W) vs time ($T_b=100$ ps) in a NRZ-OOK system consisting of 
             5 100-km
             spans of transmission fiber [CD$_{100}$=17ps/(nm$\cdot$km), $A_{eff}$=80 $\mu$m$^2$].
             Each span is followed by a 13-km
             dispersion compensation fiber [CD$_{13}$=-120ps/(nm$\cdot$km), $A_{eff}$=30 $\mu$m$^2$].
             Before the 1st 100-km fiber, there is  a precompensation of -120$\times$6 ps/nm, whereas the  
             5th 100-km fiber is followed by the compensation of -120$\times$5 ps/nm.
             The input mark power is 10 mW. Birefringence parameters are 
             $D_{PMD}$=2.0 ps/(km)$^{1/2}$, $L_{corr}=10$ m, and $\Lambda_{beat}=50$ m.
             There is very little difference between the RK4IP solution  (\ref {RK4IP_2d}) 
             and the M-PMD approx using (\ref {Manak_PMD_App}).
            }
    \label{CNLSE_Mnkv5}
\end{center}
\end{figure}

Fig. {\ref {CNLSE_Mnkv5}} shows good agreement  between 
RK4IP result and M-PMD approx for a NRZ-OOK system with 5 100-km spans of transmission fiber and
5 13-km spans of dispersion compensation fiber. 
To get the received photoelectric pulses shown in Fig. {\ref {CNLSE_Mnkv5}}, a precompensation fiber  
(6 km) is inserted before the first 100-km fiber, whereas the last  
compensation fiber is 5 km long. 
The birefringence related parameters are
$\Lambda_{beat}=50$ m, $L_{corr}$=10 m, and $D_{PMD}$=2.0 ps/(km)$^{1/2}$, which means
average DGD is around 2.0$\sqrt{5(100+13)-2}\approx$47 ps with 
the birefringence direction being randomly changed every 10 m. 
%Both results of RK4IP (dashed) and M-PMD approx (dot-dashed) are obtained using the same dispersion map.
As in {\bf \ref{RK4IP_SSFM}}, the input optical filter with bandwidth of $B=1.75B_e$ is used to generate 
rounded edges for NRZ signal.
Other paratemeters 
given in the caption of 
Fig. \ref {CNLSE_Mnkv5} are based on the discussion of Ref. \cite{Marks06}.   
As plotted in Fig. {\ref {CNLSE_Mnkv5}}, there is no significant  difference between the RK4IP solution 
(\ref {RK4IP_2d})
and the M-PMD approx (\ref {Manak_PMD_App}). 
To confirm that such  agreement is not because of the weak enough nonlinearity,
one can increase the nonlinear coefficient in (\ref{Manak_PMD_App}) from $\frac{8}{9}\gamma=1.12$/(W$\cdot$km)  
to $\frac{8}{9}\gamma=1.26$/(W$\cdot$km) for the transmission fibers and from $\frac{8}{9}\gamma=2.99$/(W$\cdot$km)
to $\frac{8}{9}\gamma=3.36$/(W$\cdot$km) for the compensation fibers. Our result shows that the discrepancy between 
the two approximations can be more than 20$\%$ (not plotted in Fig. \ref{CNLSE_Mnkv5}).

\begin{figure}
\vskip -6mm
\begin{center}
    \begin{tabular}{cc}
     %\resizebox{110mm}{!}{\includegraphics{MnkvCNLSE_RZ_OOK50spn111P2CD4.5t.4.eps}}%CNLSE/CNLSE_2
     \resizebox{110mm}{!}{\includegraphics{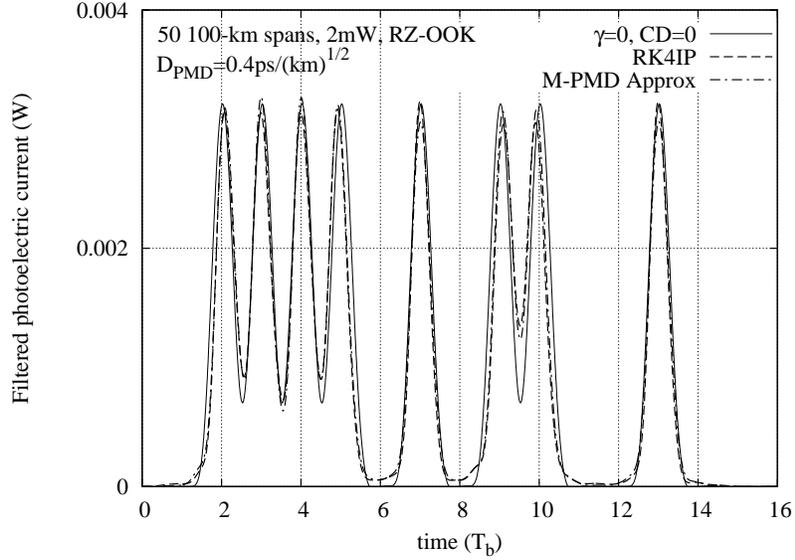}}%CNLSE/CNLSE_2
     \end{tabular}
    \vskip  -2mm
    \caption{The filtered photoelectric current (W) vs time ($T_b=100$ ps) in a RZ-OOK system using 
             50 100-km eLEAF fibers [average CD$_{100}$=4.5ps/(nm$\cdot$km), $A_{eff}$=72 $\mu$m$^2$],
             precompensation of -117 ps/nm before the first 100-km fiber, -429 ps/nm compensation per span, and
             -234 ps/nm compensation after the 50th 100-km fiber.
             The input mark power is 2 mW and the PMD coefficient is $D_{PMD}$=0.40 ps/(km)$^{1/2}$.
            }
    \label{CNLSE_Mnkv50_4.5}
\end{center}
\end{figure}

\begin{figure}
\vskip -6mm
\begin{center}
    \begin{tabular}{cc}
     %\resizebox{110mm}{!}{\includegraphics{MnkvCNLSE_RZ_OOK50spnP2CD6t.41.eps}}%CNLSE/CNLSE_2
     \resizebox{110mm}{!}{\includegraphics{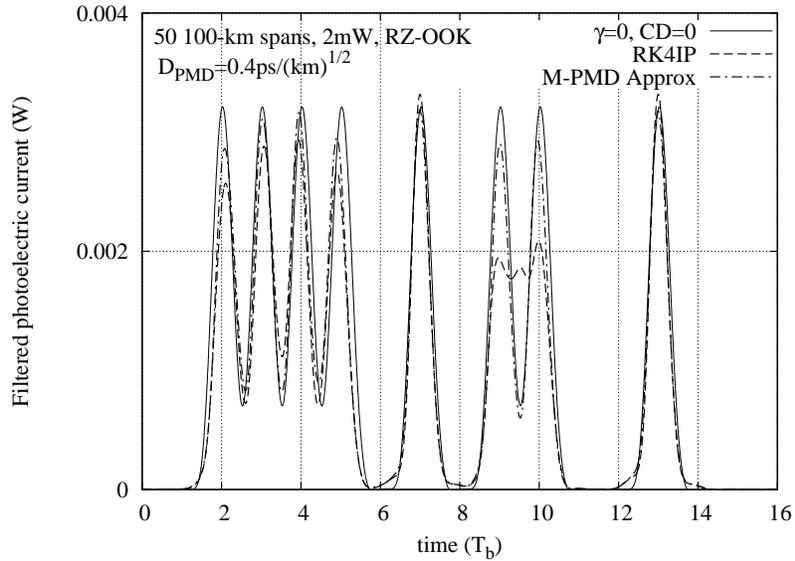}}%CNLSE/CNLSE_2
     \end{tabular}
    \vskip  -2mm
    \caption{Same as Fig. \ref{CNLSE_Mnkv50_4.5}, except that
             CD$_{100}$=6.0ps/(nm$\cdot$km).
             Because of the change of CD,
             each span is followed by a -585 ps/nm dispersion compensation.
             The last 100-km fiber is followed by the compensation of -390 ps/nm.
             %Changing $D_{PMD}$ from  0.41 ps/(km)$^{1/2}$ to
             %0.40 ps/(km)$^{1/2}$ does not significantly change the results plotted here. 
             %This is
             %different from those shown in Fig.\ref {CNLSE5b}, where the transmission line is
             %less than
             %570 km and its $D_{PMD}$ is $\sim 5 $ ps/(km)$^{1/2}$.
            }
    \label{CNLSE_Mnkv50_6}
\end{center}
\end{figure}

Fig. \ref {CNLSE_Mnkv50_4.5} shows the RK4IP solution and M-PMD approx for a
RZ-OOK system using 50 100-km spans of eLEAF fiber \cite{Wiesenfeld05}, 
with
effective mode area being   $A_{eff}$=72 $\mu$m$^2$ and
CD in the range of $3\sim6$ ps/(km$\cdot$nm). To get the received photoelectric pulse shown in
Fig. \ref {CNLSE_Mnkv50_4.5}, each 100-km fiber with CD$_{100}$=4.5ps/(km$\cdot$nm) is followed
by -429 ps/nm compensation.
Also, precompensation of %-78 
-117 ps/nm before the first 100-km fiber and last
compensation of %-273 
-234 ps/nm after the 50th 100-km fiber are used.
The birefringence related parameters are   $L_{corr}=100$ m and $\Lambda_{beat}=50$ m. 
%Other parameters in the caption of Fig. \ref {CNLSE_Mnkv50_4.5} are based on the discussions in Refs.
%\cite{Shtaif06,Wiesenfeld05}.
As shown in Fig. \ref {CNLSE_Mnkv50_4.5}, the agreement between the RK4IP solution and M-PMD approx is not as good as 
that of Fig. \ref {CNLSE_Mnkv5}. 
Considering that, due to the linear-nonlinear interaction accumulated in such 50 100-km fibers, any small change in the
related parameters  will lead to significant change
in received pulse shape, the M-PMD approx (dot-dashed) in Fig. \ref {CNLSE_Mnkv50_4.5} can be viewed as 
a reasonably good approximation of the CNLSE solution obtained using RK4IP (dashed).

Fig. \ref {CNLSE_Mnkv50_6} shows that, when the CD parameter of eLEAF fiber is increased to its worst case 
[6.0ps/(km$\cdot$nm)],
the discrepancy between RK4IP and M-PMD approx becomes larger than that of 
Fig. \ref {CNLSE_Mnkv50_4.5}, due to the increased interaction between linear and nonlinear terms in the CNLSE.
%Details of the dispersion map for this case is given in the caption of Fig. \ref {CNLSE_Mnkv50_6}.

%%%%%%%%%%%%%%%%%%%%%%%%%%%%%%%%%%

\section{Summary}

Based on the RK4IP method for 1D NLSE \cite{Hult07} as well as the analytical solution for the linear terms in CNLSE 
\cite{Marcuse97}, RK4IP method for 2D CNLSE is presented in this work.
Without rotating the coordinate system for each step, our approach can be applied to 
a fiber with general form of 
birefringence. Besides,  the Kerr nonlinearity in the CNLSE is treated without approximation.
%the condition for M-PMD approx is not required in our approach.
Since there is no split-step approximation for each step and the local error method of Ref. \cite{Sinkin03} is used,
for normal fibers with random birefringence, 
the step size using RK4IP can be orders of magnitude larger than
$\Lambda_{beat}$ and $L_{corr}$, depending on the intensity of the linear-nonlinear interaction. 
Without birefringence effect, the RK4IP step size   
can be increased to the same order of magnitude as $L_{D}$  and/or $L_{N}$, 
or, around $6\sim10$ times the step size using the approach of Ref. \cite{Marcuse97}. 

In the parameter regime where communication fibers normally operate, 
our results are consistent well with the results using M-PMD approx (\ref {Manak_PMD_App}) 
\cite{Wai96,Marcuse97,Marks06}. 
Increased interaction between the
linear and nonlinear terms  in the CNLSE will lead to increased  
discrepancy between RK4IP solution and M-PMD approx.

%%%%%%%%%%%%%%%%%%%%%%%%%%%%%%
\vskip 0.10 in
\noindent
{\bf Acknowledgment}

\vskip 0.10 in

\noindent
The authors acknowledge the financial support from Canadian funding agencies: NSERC. 
\vskip 0.20 in
\noindent
%%%%%%%%%%%%%%%%%%%%%%%%%%%%%%%%%%%%%%%%%%%%%%%%%%%%%
%%%%%%%%%%%%%%%%%%%%%%%%%%%%%%%%%%%%%%%%%%%%%%%%%%%
%%%%%%%%%%%%%%%%%%%%%%%%%%%%%%%%%%%%%%%%%%%%%%%%%%%%
%%%%%%%%%%%%%%%%%%%%%%%%%%%%%%%%%%%%%%%%%%%%%%%%%%%%%
{\bf Appendix: Optical field in the fiber without nonlinearity}
%%%%%%%%%%%%%%%%%%%%%%%%%%%%%%%%%%
%%%%%%%%%%BBBBBBBBBBBBBBBBB%%%%%%%%%
\vskip 0.20 in
\noindent
For a system with $\gamma=0$, Eq. (\ref {CNLS_3f}) can be reduced to
\begin{eqnarray}
j\frac{\partial }{\partial z}|u(z,t)\rangle
+\vec \beta_0(z)\cdot\vec \sigma\big(\Delta\beta \!+\!j\Delta\beta_{\omega}\frac{\partial }{\partial t}\big) |u(z,t)\rangle
\!-\!\frac{\beta_{\omega\omega}}{2}\frac{\partial ^2}{\partial t^2}|u(z,t)\rangle=0
\label{CNLS_linear}
\end{eqnarray}
In frequency domain, Eq. (\ref {CNLS_linear}) can be simplified as 
\begin{eqnarray}
j\frac{\partial }{\partial z}|U(z,\omega)\rangle
+\Big[\vec \beta_0(z)\cdot\vec \sigma\big(\Delta\beta \!-\!\Delta\beta_{\omega}\omega\big)
\!+\!\frac{\beta_{\omega\omega}}{2}\omega^2\Big]|U(z,\omega)\rangle=0,
\label{CNLS_linear2}
\end{eqnarray}
where $|U(z,\omega)\rangle=\int |u(z,t)\rangle e^{-j\omega t}dt$.
(Note that in Ref. \cite{Marcuse97}, the Fourier transformation  was defined to be   
$|U(z,\omega)\rangle=\int |u(z,t)\rangle e^{j\omega t}dt$.) 

Assuming
\begin{equation} 
|U(z,\omega)\rangle
=M(z,\omega)e^{j\frac{\beta_{\omega\omega}}{2}\omega^2(z-z_0)}|U(z_0,\omega)\rangle
\equiv I(z,z_0) |U(z_0,\omega)\rangle,
\label{trans}
\end{equation}
Eq. (\ref {CNLS_linear2}) is equivalent to  
the following differential equation for the $2\times 2$ matrix $M$ \cite{Marcuse97}:
\begin{equation}
j\frac{\partial M(z,\omega)}{\partial z}+\vec \beta_0(z)\cdot\vec \sigma\big(\Delta\beta \!-\!\Delta\beta_{\omega}\omega\big)M(z,\omega)=0.
\label{C_linearM}
\end{equation}
The correct solution of (\ref {C_linearM}) should be in the form of \cite{Marcuse97} 
\begin{eqnarray}
%M(z,\omega)=\prod_{i=1}^{N_h} m_i=\prod_{i=1}^{N_h} e^{j\delta_i \eta\big(\vec \beta_0(z_i)\cdot\vec \sigma\big)}
%=e^{jB\vec \beta_0\cdot \vec \sigma},
&&
M\!(\!z,\omega\!)\!=\!
m_{N_h}\!\cdot m_{N_h-1}\cdots m_2\cdot m_1 %=>e^{jB\vec \beta_0\cdot \vec \sigma},\;\;\; 
\label{M_solutn}
\\
&&
m_i\!=\!e^{j\delta_i \eta \big(\vec \beta_0(z_i)\cdot\vec \sigma\big)}
\;\;\;\;(i\!=\!1,\cdots,N_h,\;\;\eta\!\equiv\! \Delta\beta \!-\!\Delta\beta_{\omega}\omega)
\label{M_solutn2}
\end{eqnarray}
where %$\vec \beta_0=\vec B/B$ ($B=|\vec B|$) and  
$\delta_i=z_{i-1}-z_i$. 
%Note that $z_i$ is not necessary larger
%than $z_{i-1}$. 
When $z_i>z_{i-1}$, $\delta_i$ is the length of the $i$th plate with its birefringence direction being  
represented by $\vec \beta_0(z_i)$. 
In (\ref {M_solutn}), 
$N_h$ is the number of the plates between $ z_0$ and $z=z_0+h$ and the index $N_h$ denotes the $N_h$th  plate.
The expression of $M(z,\omega)$ given by (\ref {M_solutn}) can also be extended to
the case of $z-z_0<0$, provided to notice that 1) for each plate, $\delta_i<0$;
2) $m_i$ 
($i=1,\cdots,N_h$) are ordered from $z_0$ to $z$, i.e., $m_1$ is the first left plate of $z_0$.
[ In (\ref {M_solutn2}) $\eta$ is determined by the strengths  of the phase birefringence ($\Delta\beta$) and the 
group birefringence ($\Delta\beta_{\omega}$),
which is assumed to be
independent of $z$ in this work.]

Using a fixed coordinate system in our approach,   
the result of (\ref {M_solutn}) can be simply obtained by
algebra operations such as  \cite{Damask05} 
\begin{eqnarray}
\!\!\!\!\!\!&\!\!\!\!\!\!&\!\!\!\!\!\!
m_1=\cos(\delta_1 \eta)+j\sin(\delta_1 \eta)\vec \beta_0(z_1)\cdot\vec \sigma\equiv \mu_0^{(1)}+j\vec\mu^{(1)}\cdot\sigma,
\nonumber \\
\!\!\!\!\!\!&\!\!\!\!\!\!&\!\!\!\!\!\!
m_1m_2\!=\!\big[\mu_0^{\!(\!1\!)}\mu_0^{\!(\!2\!)}\!-\! \vec\mu^{(\!1\!)}\!\cdot\!\vec\mu^{(\!2\!)}\!\big]
\!+\!j\vec \sigma\!\cdot\!\big[\mu_0^{\!(\!1\!)}\vec\mu^{(\!2\!)}\!
+\!\mu_0^{\!(\!2\!)}\vec\mu^{(\!1\!)}\!-\!(\vec\mu^{(\!1\!)}\!\times\!\vec\mu^{(\!2\!)}) \big]
\equiv \mu_0^{\!(\!12\!)}+j\vec\sigma\cdot\vec \mu^{(\!12\!)}.
\label{basics}
\end{eqnarray}
%The Hermitian of $M$ is $M^\dagger\!=\!m_1^\dagger\cdot m_2^\dagger\cdots m_{N_h-1}^\dagger\cdot m_{N_h}^\dagger$
%with $m_i^\dagger\!=m_i^{-1}=\!e^{-j\delta_i \eta\big(\vec \beta_0(z_i)\cdot\vec \sigma\big)}$.

Obviously, in time domain,
the transform matrices in (\ref {trans}) have the form of ($i=1,\cdots,N_h$) 
\begin{eqnarray}
\!\!\!\!&\!\!\!\!&\!\!\!\!
\hat {I}(z,z_0)=\hat M(z,t)\hat{d}(z-z_0)
\nonumber\\
\!\!\!\!&\!\!\!\!&\!\!\!\!
\hat{d}(z-z_0)|_{z=z_0+h}=
e^{-jh\frac{\beta_{\omega\omega}}{2}\frac{\partial ^2}{\partial t^2}}
\nonumber\\
\!\!\!\!&\!\!\!\!&\!\!\!\!
\hat M(z,t)=\hat{m}_{N_h}\cdot \hat{m}_{N_h-1}\cdots\hat{m}_{2}\cdot\hat{m}_1,\;\;\;\;\;
\hat {m}_i=e^{j\delta_i \big( \Delta\beta \!+\!j\!\Delta\beta_{\omega}\frac{\partial}{\partial t}\big)\big(\vec \beta_0(z_i)\cdot\vec \sigma\big)}.
\label{trans_t}
\end{eqnarray}

%It seems that  the solution of (\ref {C_linearM}) should be
%\begin{equation}
%M=e^{j\int_{z_0}^z dz'\big(\vec \beta_0(z')\cdot\vec \sigma\big)\eta}.
%\label{M_approx}
%\end{equation}
%However, according to Baker-Hausdorff formula, this is only the zero order approximation of (\ref {M_solutn}),  
%since the commuting operator 
%$[\int_{z_0}^{z}dz' \vec \beta_0(z)\cdot\vec \sigma, \int_{z}^{z+\Delta z}dz' \vec \beta_0(z)\cdot\vec \sigma]$  is not 
%proportional to the unit matrix.

\end{document}